# Archimedean solid-like superconducting framework in phase-separated $K_{0.8}Fe_{1.6+x}Se_2$ $(0 \leq x \leq 0.15)$

Z. Wang, Y. Cai, Z. W. Wang, C. Ma, Z. Chen, H. X. Yang, H. F. Tian, and J. Q. Li*

*Beijing National Laboratory for Condensed Matter Physics, Institute of Physics, Chinese Academy of Sciences, Beijing 100190, China*

The superconducting (SC) phase in the phase-separated (PS) $K_{0.8}Fe_{1.6+x}Se_2$ $(0 \leq x \leq 0.15)$ materials is found to crystallize on Archimedean solid-like frameworks, this structural feature originate from a spinodal phase separation (SPS) at around $T_s \approx 540K$ depending slightly on the Fe concentration. Two stable phases in $K_{0.8}Fe_{1.6+x}Se_2$ are demonstrated to be the SC $K_{0.5}Fe_2Se_2$ and antiferromagnetic (AFM) $K_{0.8}Fe_{1.6}Se_2$. The spinodal waves go along the systematic [113] direction and result in notable lamellar structure as illustrated by using the strain-field theoretical simulation. The 3-dimentional SC framework is constructed by hollow truncated octahedra similar with what discussed for Archimedean solids. Based on this structural model, we can efficiently calculate the volume fraction of SC phase in this type of PS SC materials.

Phase separation and structural inhomogeneity as critical structural issues have been extensively investigated in a variety of strongly correlated systems, such as the high-$T_c$ cuprate superconductors [1-3] and the colossal magneto-resistance manganese [4, 5]. Furthermore, it is often noted that the coexistence/competition between SC and magnetic states could not only lead to structural inhomogeneities but also yield debated issues for understanding of the SC mechanism in the high-Tc materials. Recent study of the alkali-metal intercalated FeSe superconductors $A_yFe_{2-x}Se_2$ (A = K, Rb, Tl and Cs) revealed notable microstructure features in correlation with structural phase separation [6-9], notable structural and physical properties have been analyzed based on the coexistence of the insulating AFM $K_2Fe_4Se_5$ phase and SC phase. In order to understand the phase separation behaviors, microstructure properties of well-characterized $A_yFe_{2-x}Se_2$ materials have been examined by using transmission electron microscopy (TEM) [10, 11], X-ray diffraction (XRD) [12] and scanning tunneling microscopy (STM) [13]. Measurements of Mössbauer spectrum [14] and nuclear magnetic resonance (NMR) [15] on the SC samples suggest that only 10-20% volume fraction really exhibits superconductivity and the rest 80-90% of the volume remains antiferromagnetic. Moreover, the SC property of this PS system could be tuned by the thermal annealing process [16, 17], therefore, understanding the dynamic process of phase transition is crucial for elucidating the underlying formation mechanism of the PS state in this system. Due to the complexity of the PS phenomena in present system, there is still ongoing debate about the PS nature and atomic structure of the SC phase. In this paper, we will report on direct observations of phase separation at high temperatures as revealed by *in-situ* TEM observations, illustrating the formation of SC framework embedded in the insulating AFM $K_2Fe_4Se_5$ phase. It is well demonstrated that the SC phase and related domain structure are actually arising from a SPS at high temperatures. The 3-dimensional SC framework is made up of hollow truncated octahedral components similar with what discussed for the well-known Archimedean solids [18, 19]. The atomic structure of the SC phase has been determined by using aberration-corrected TEM and scanning transmission electron microscopy.

A series of well-characterized SC samples, both single-crystal and polycrystalline materials, were used in the present study. In particular, materials with nominal composition of $K_{0.8}Fe_{1.6+x}Se_2$ ($0 \leq x \leq 0.2$) have been extensively investigated to understand the relationship of microstructural and physical properties for the PS states. High-resolution TEM observations and *in-situ* microstructure analysis were performed on a Tecnai-F20 transmission electron microscope with a heating holder. Aberration-corrected STEM investigations were performed by using a JEOL ARM200F equipped with double aberration correctors and cold field emission gun with imaging resolution of 78 picometers. A few specimens for microstructure analysis have been prepared by using focused ion beam (FIB) for revealing the top and side views of the lamellar structures in single crystalline samples.

We firstly focus on the high-temperature structural transitions in correlation with the PS state and microstructural inhomogeneity that were frequently observed in $K_{0.8}Fe_{1.6+x}Se_2$ superconductors. *In-situ* TEM investigations revealed that visible changes of microstructure, such as domain patterns and ordered states, appear mostly in the temperature range from 570K to 450K which depends slightly on the nominal chemical compositions. One of the most striking features is the appearance of micro-domains in all crystals following with lowering temperature, and these domains in general can be clearly seen as regular stripes on the **a-b** plane as briefly discussed in our preview publication [11]. Figure 1(a-c) show the *in-situ* STEM high-angle annular dark field (HAADF) images and corresponding electron diffraction patterns to illustrate the structural transitions obtained from a SC crystal of $K_{0.8}Fe_{1.7}Se_2$ ($T_c$ = 32K). In fact, our observations revealed that all $K_{0.8}Fe_{1.6+x}Se_2$ (0 < x < 0.2) samples in general have a high-temperature homogeneous phases in which Fe-vacancies are disordered within a tetragonal structure, as recognizable by electron diffraction pattern and STEM image obtained above $T_s$ = 540K in Fig. 1(a), and no any superstructure or micro-domains can be observed. It is demonstrated that stripe-like domains following with the structural transition at $T_s$ commonly appear in present materials as illustrated in the STEM image of Fig. 1(b), electron diffraction pattern obtained from the areas with dark contrast suggests the formation of the known superstructure with $\mathbf{q_1}$ = 1/5(3, 1, 0) induced by Fe-vacancy order. On the other hand, these bright domains in the STEM image contain relatively rich Fe and poor K as clearly indicated by the line-scanning EDX results in the inserted image of Fig. 1(b). Careful analysis of the experimental data suggests structural transformation at $T_s$ = 540K can be fundamentally understood as a SPS process, which results in two visible structural alterations, i.e. the fluctuations of Fe/K concentration yielding domain structures and the Fe-vacancy order in matrix, then the crystallization of SC phase within these Fe-rich domains is below the spinodal transition and associated with a K-vacancy ordering characterized by a modulation of $\mathbf{q_2}$ = 1/2(1, 1, 0) as shown in Fig. 1(c). The K-ordering transition in general occurs at visibly lower temperature than $T_s$ and yields recognizable nano-scale phase separation in Fe-rich stripes, e.g. the $K_{0.8}Fe_{1.7}Se_2$ superconductor showing the K-ordering transition at about $T_o$ = 480K as directly revealed in our *in-situ* TEM observations. Moreover, the reversibility of these high-temperature phase transitions has been checked by *in-situ* heating-cooling cycles using XRD and TEM in a few well-characterized samples as shown in Fig. S1 and S2. Though hysteresis features can been seen for the transition temperatures, the structural transitions in present SC system are fundamentally reversible as similarly discussed for the SPS in other materials [20].

Furthermore, the apparent effects of Fe content on the SPS process and domain structures have been also observed in $K_{0.8}Fe_{1.6+x}Se_2$ system. Figure 2(a) shows a brief phase diagram for the $K_{0.8}Fe_{1.6+x}Se_2$ system, in which the SPS and K-ordering transition are clearly indicated and the critical temperatures ($T_s$ and $T_o$) decrease slowly with the increase of Fe content. For instance, the sample with x = 0 shows $T_s$ = 578K [21], and $T_s$ is about 540K for sample with x = 0.15 [22]. It is suggested that in the SPS process the volume fraction of SC phase could range from 5% to 40% depending on the Fe-concentration. We herein employ a 2-dimensional Monte Carlo method using the conventional phase-field approach to simulate the microstructure evolution in the SPS process in basic a-b plane, and the detail of this method is presented in supplementary material. Figure 2(b) shows the experimental and simulated images for a number of samples with different Fe-concentrations, the theoretical data are in good agreement with the experimental ones. It is noted that the increase of Fe content (x) in the $K_{0.8}Fe_{1.6+x}Se_2$ system commonly yields continuous changes of the Fe-rich domains from micro-dots to stripes implanted in the matrix, which plays an important role for the appearance of superconductivity and formation of SC frameworks in this kind of PS materials. Moreover, as commonly discussed in other SPS materials, $K_{0.8}Fe_{1.75}Se_2$ system follows with spinodal process toward thermodynamic equilibrium depending on both temperature and time [23]. Actually, our structural investigations also revealed that the PS nature and domain patterns in the SC materials could change visibly with the thermal process in sample preparations, which also lead to the changes of the transport property of these samples (see Fig. S4). For instance, samples annealed for different time at a temperature above $T_s$ often shows up a number of composition fluctuations with different wavelengths between a few nanometers and micrometers. A few SEM images of $K_{0.8}Fe_{1.75}Se_2$ annealed at 550K for different time are presented in supplementary materials of Fig. S3, the theoretical simulated patterns are also shown for comparison.

The Fe-rich domains in the SC materials actually play an important role for the appearance of superconductivity in $K_{0.8}Fe_{1.6+x}Se_2$, therefore, the three-dimensional configuration of the domain structures as an important issue has been also concerned in our study. In our TEM and SEM observations, the specimens have been cut for different orientations, and images along the relevant zone-axis directions have been obtained to reveal the top and side views of the domain structures. Figure 3(a-c) show the experimental data obtained from SEM observations along [100], [110] and [001] zone-axis directions, illustrating the complex domain structures in $K_{0.8}Fe_{1.7}Se_2$ on the relevant crystal planes. It is concluded that, in association with the SPS, the spinodal fluctuation wave for Fe-concentration goes along the systematic [113] direction and crystallographically equivalent variants, as a result, the Fe-rich domains form a 3-dimensional framework running through the $K_{0.8}Fe_{1.7}Se_2$ materials. Figure 3(e) shows an ideal 3-dimensional schematic diagram for Fe-rich SC phase embedded in the AFM $K_{0.8}Fe_{1.6}Se_2$ phase. The Fe-rich framework consists of notable octahedral components with their sizes of around a few microns as shown in the inset of Fig. 3(d). The angle-resolved electrical transport measurements on $K_{0.8}Fe_{1.6+x}Se_2$ superconductor [24] suggest that fourfold symmetric behaviors of resistivity out of plane is closely related with the SC stripes, which also evident our 3-dimensional model of SC framework. Actually, the local microstructure features of this framework depend on the Fe content and thermal conditions for sample preparation as discussed above. In geometric pint of view, this structural model is similar with the Archimedean solid discussed for a space-filling tessellation in a 3-dimensinal Euclidean space made up of truncated octahedra [19, 25]. Based on this structural model, the volume fraction of the SC phase could be estimated based on the measurements of the relative size of structural domains as shown in SEM or TEM images. For instance, the volume fraction of SC phase for the $K_{0.8}Fe_{1.7}Se_2$ superconductor could be calculated to be about 17%, which is in good agreement with the experimental data obtained from Mössbauer spectrum [14] and nuclear magnetic resonance (NMR) [15]. The detailed calculation of the SC fraction based on the Archimedean solid model is illustrated in supplementary materials.

It is also noted in our *in-situ* TEM observations, though the SPS could result in the Fe-rich domains embedded in $K_{0.8}Fe_{1.6+x}Se_2$ materials, the nucleation and formation of SC phase in the Fe-rich domains is likely characterized by the gradual emergence of an ordered state of $q_2 = 1/2(1, 1, 0)$ below $T_s$. Moreover, the formations of the $q_2$ SC phase in the Fe-rich domains often follow with microstructural changes and PS on nano-scales. As a result, we can often see certain interesting nano-scale PS patterns in the Fe-rich domains, such as stripes, chessboard and layered structures as discussed in our previous publications [10, 25] and in Fig. S5.

It is commonly noted that $K_{0.8}Fe_{1.6+x}Se_2$ ($0.05 \leq x \leq 0.15$) superconductors often contain well-crystallized $\mathbf{q_2}$ phase within the Fe-rich domains, we therefore chose the $K_{0.8}Fe_{1.7}Se_2$ superconductor as the typical sample to indentify the crystal structure of SC phase in present study. Though our structural and chemical analysis cannot obtain an accurate composition for the $\mathbf{q_2}$ phase in a specific area, we actually take the average data from a number of the Fe-rich domains for experimental analyses. It is demonstrated that the average composition of Fe-rich stripes can be roughly written as $K_yFe_2Se_2$ with $y \approx 0.5$. As shown in Fig. 4(a), our structural measurements in the high-resolution TEM images revealed that $\mathbf{q_2}$ phase has a relatively longer $\mathbf{c}$-axis (~2%) in comparison with the insulating AFM $K_2Fe_4Se_5$ phase, and this result is fundamentally in agreement with our pervious x-ray diffraction data [11]. Figure 4(b) shows a Cs-STEM HADDF image of $\mathbf{a\text{-}b}$ plane for the SC phase, in which the atom columns show bright contrast, and their intensities depend on atomic number and occupation ratio for each column. It is clearly demonstrated that the intensity of K/Se columns alternates along [110] direction, which is considered to be induced by the present of a K-vacancy ordering as shown in the structural model in Fig. 4(c). This superstructure has a pseudo-tetragonal symmetry with lattice parameters of $a \approx b = 5.53$ Å and $c = 14.19$ Å. The simulated electron diffraction patterns and HADDF image are in good agreement with the experimental ones as shown in Fig. 4(d) and the inserted image of Fig. 4(b).

It is known that spinodal decomposition is a significant mechanism for understanding the dynamic mixture of liquids or solids [26, 27]. Our structural analyses suggest that the rich microstructure phonoma and superconductivity in $K_{0.8}Fe_{1.6+x}Se_2$ materials are essentially in correlation with the composition modulation of K/Fe cations on micron-scale. The spinodal wave shows up visible dynamic features at around the critical temperature ($T_s$) so that microstructure properties and superconductivity could depend visibly on thermal conditions for samples preparations [16, 17]. It is also noted that a small fraction of element substitution for Fe atoms in present system often yield visible changes on SPS and microstructure properties [28]. As the typical results, the SC stripes are found to adopt a variety of different patterns in $K_{0.8}Fe_{1.75-x}M_xSe_2$ (M=Mn, Zn, Ni…) materials.

In summary, $K_{0.8}Fe_{1.6+x}Se_2$ SC materials show a rich variety of remarkable microstructure phenomena. Structural analyses reveal that this kind of materials in general contains a SC framework embedded in the AFM $K_{0.8}Fe_{1.6}Se_2$ matrix. Our aberration-corrected TEM/STEM studies show the SC phase in $K_{0.8}Fe_{1.6+x}Se_2$ can be written as $K_{0.5}Fe_2Se_2$. Experimental observations and theoretical simulations demonstrate that this notable structural feature results from a SPS following by crystallization of the SC phase with a K-order state. The 3-dimensional SC frameworks is made up of hollow (stretched) truncated octahedra with their square faces linked together similar with what discussed for the well-known Archimedean solids. Based on these structural features, the SC fraction in the PS $K_{0.8}Fe_{1.6+x}Se_2$ materials can be efficiently estimated, therefore it is now possible to examine the detailed influence of the thermal conditions and Fe-content on superconductivity in this kind of PS superconductors.

Acknowledgement

This work was supported by National Basic Research Program of China 973 Program (Grant Nos. 2011CBA00101, 2010CB923002, 2011CB921703, 2012CB821404), the Natural Science Foundation of China (Grant Nos. 11274368, 51272277, 11074292, 11004229, 11190022), and Chinese Academy of Sciences.

**Figure caption:**

Fig. 1 *In-situ* STEM images and corresponding electron diffraction patterns of **a-b** plane at different temperatures for $K_{0.8}Fe_{1.7}Se_2$ illustrate the formation of PS state. (a) STEM image and DP with no superstructure spots above $T_s$. (b) STEM image with stripe patterns along [110] and [1-10] directions, the EDX line-scanning data of Fe and K elements are shown in the inserted image. DPs taken from Fe-poor matrix (area A) with modulation of $\mathbf{q_1}$ = 1/5(3, 1, 0) and the Fe-rich domain (area B) below $T_s$. (c) DPs taken from the Fe-rich domains (area B) with modulations of $\mathbf{q_2}$ = 1/2(1, 1, 0) and $\mathbf{q_1}$ = 1/5(3, 1, 0) below $T_o$.

Fig. 2 (a) A brief phase diagram for $K_{0.8}Fe_{1.6+x}Se_2$ SC system, the blue color scale illustrates the percentage of $\mathbf{q_2}$ phase in samples which exhibits superconductivity for x ≥ 16%. (b) Theoretical simulation and experimental patterns for $K_{0.8}Fe_{1.6+x}Se_2$, it is recognizable that the SPS depends visibly on the Fe content. In the simulated images, white regions represent for Fe-rich domains and black regions for Fe-poor matrix.

Fig. 3 (a-c) SEM images for a $K_{0.8}Fe_{1.7}Se_2$ sample illustrating the stripe patterns on (001), (010) and (110) planes, respectively. (e) 3-dimensional schematic model for SC skeleton in $K_{0.8}Fe_{1.7}Se_2$: the blue region represents the Fe-rich domains (i.e. SC $\mathbf{q_2}$ phase), and the yellow region represents the AFM $\mathbf{q_1}$ phase. The framework is composed of interconnected hollow octahedral components shown in (d) with their sizes of around a few microns.

Fig.4 (a) High-resolution image taken along [310] zone-axis direction for $K_{0.8}Fe_{1.7}Se_2$ sample, showing the interface structures between $\mathbf{q_1}$ and $\mathbf{q_2}$ phase. (b) The aberration-corrected STEM image of **a-b** plane for the SC $\mathbf{q_2}$ phase. Inserted figure is a theoretical simulated image based on the structural model in fig. 4(c), and microphotometric density curve shows the regular variation of the contrast from K/Se columns. (c) A structural model for $K_{0.5}Fe_2Se_2$ with K-ordering projected on the **a-b** plane. (d) The experimental (left) and simulated (right) DPs along [001] zone-axis direction.

Fig. 1

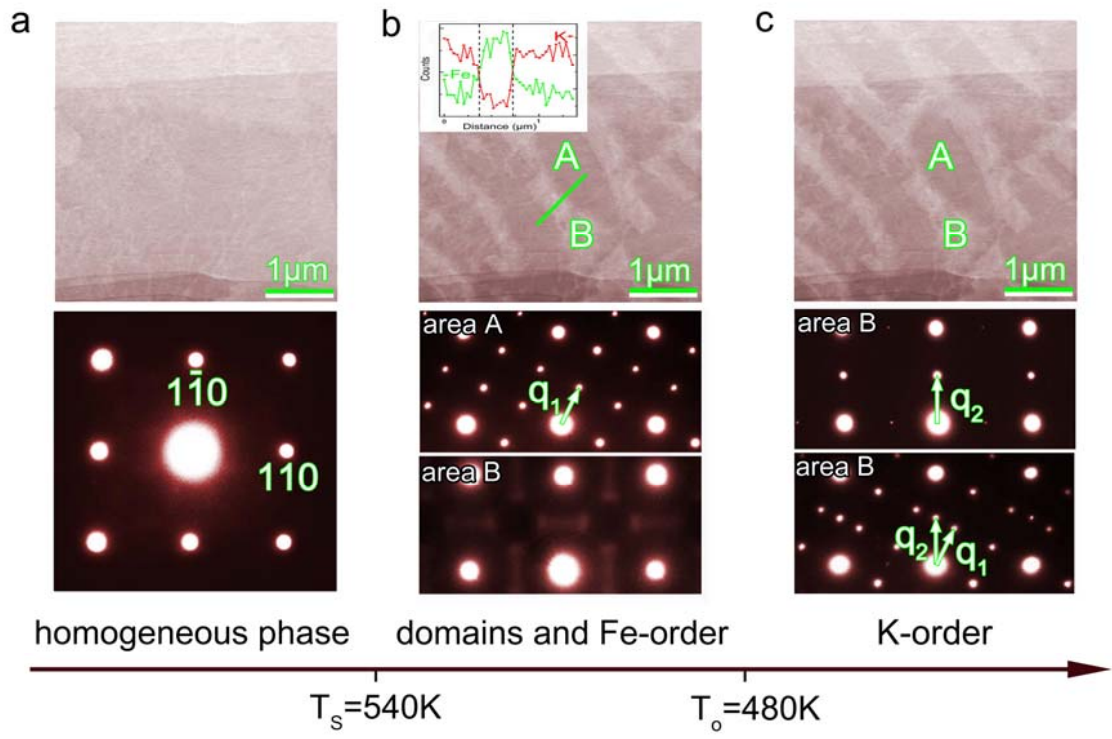

Fig.2

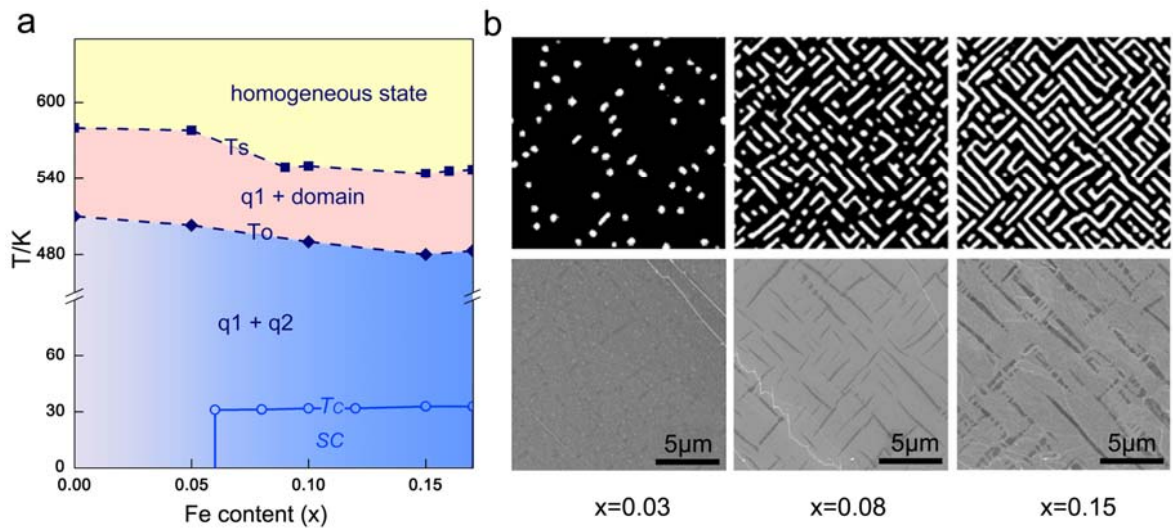

Fig.3

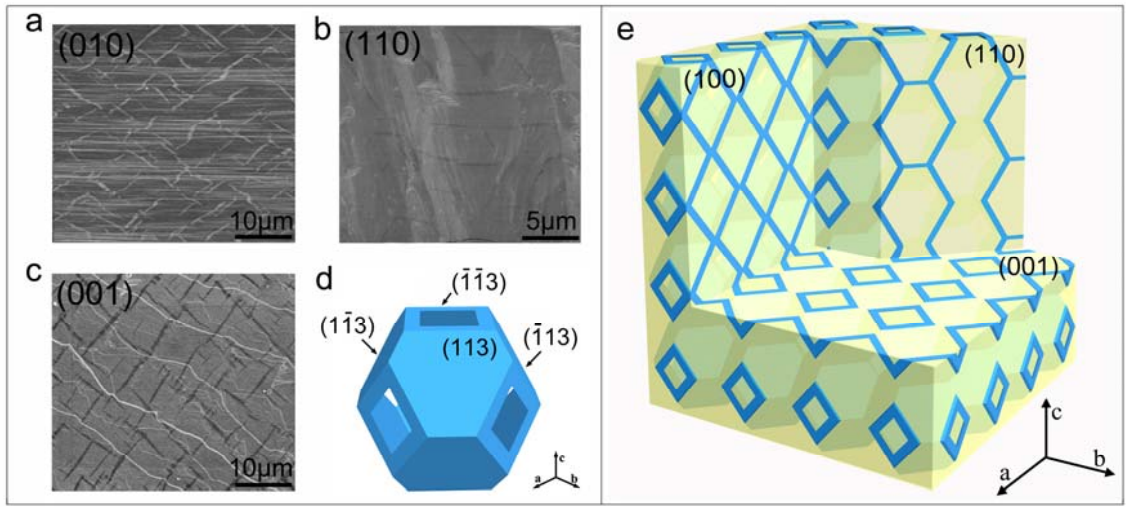

Fig.4

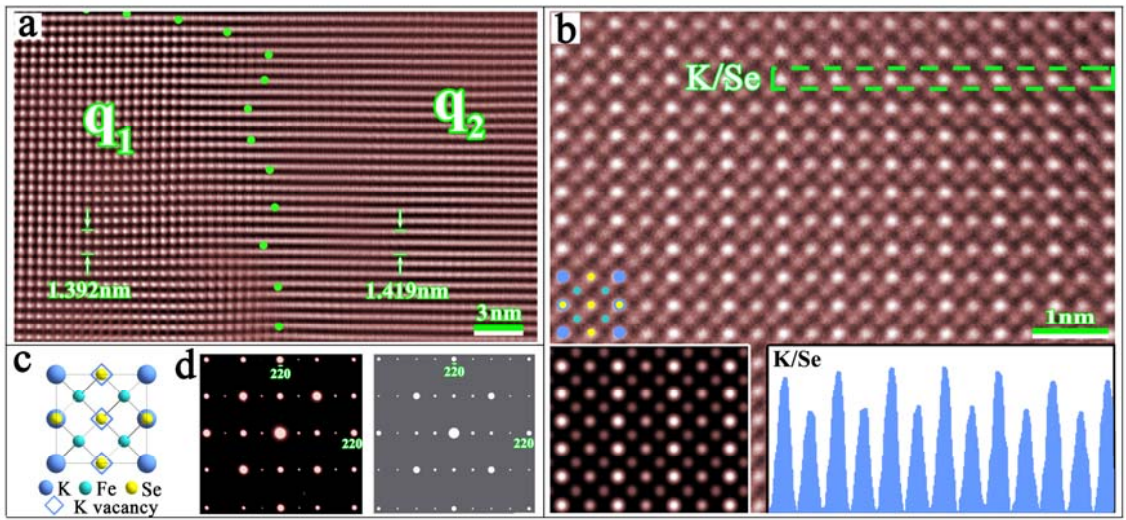

**Supplementary material for "Superconducting micro-skeleton from spinodal phase separation in $K_{0.8}Fe_{1.6+x}Se_2$ ($0 \leq x \leq 0.15$)"**

Z. Wang, Y. Cai, Z. W. Wang, C. Ma, Z. Chen, H. X. Yang, H. F. Tian, and J. Q. Li*

*Beijing National Laboratory for Condensed Matter Physics, Institute of Physics, Chinese Academy of Sciences, Beijing 100190, China*

## I. Reversibility of high temperature phase transitions in $K_{0.8}Fe_{1.6+x}Se_2$ system

In-situ X-ray diffraction experiments on single crystals and polycrystals demonstrate that $q_1$ and $q_2$ phases vanish with the increase of temperature and reform after the SPS in the cooling run, and similar experimental results were also reported in the Ref. [1]. The integral peak intensity of $q_2$ phase decreases gradually above 400K in the heating run and disappears at 490K, while it shows up gradually in the cooling run below 520K. The intensity ratio of $q_1$ and $q_2$ peak reduced after the heating and cooling circle. Careful analysis on the experiments indicates that solid solution state above $T_s$ is inclined to separate out iron grains in the $Ar_2$ environments, which results in the increase of $T_s$ and the decrease of $q_2$ phase content. The similar reversible transitions were also observed in our *in-situ* TEM experiments for $K_{0.8}Fe_{1.75}Se_2$ sample as shown in Fig. S2.

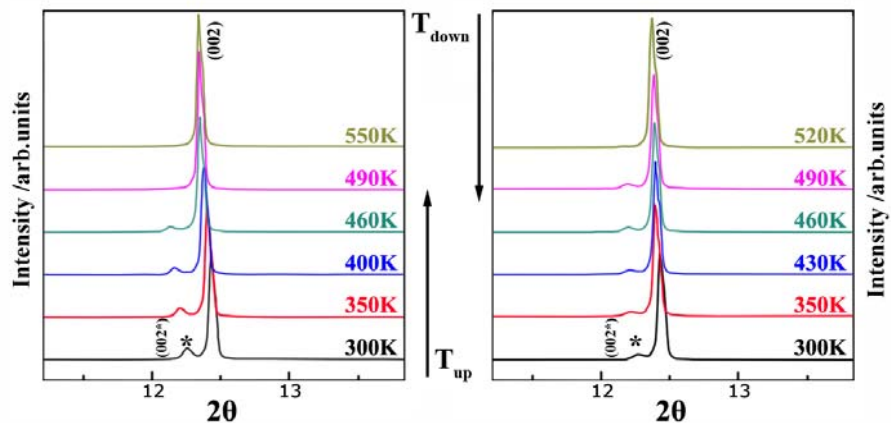

**Fig. S1** *In-situ* X-ray diffraction patterns for $K_{0.8}Fe_{1.7}Se_2$ single crystal of heating and cooling cycles. The black star indicates the peak of $q_2$ phase.

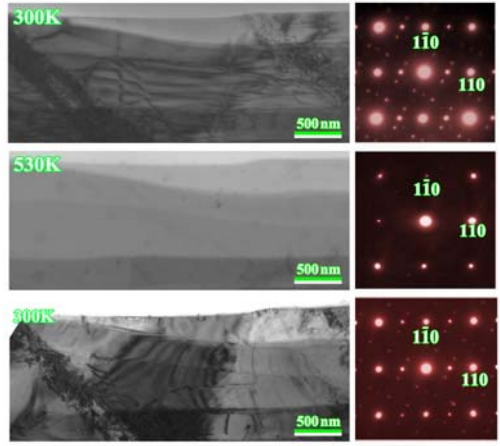

**Fig. S2** The *in-situ* TEM images of **a-b** plane and corresponding electron diffraction patterns for $K_{0.8}Fe_{1.75}Se_2$ SC samples. (a) The stripes in TEM image and two sets of superstructure spots with modulated vector $\mathbf{q_1}$ and $\mathbf{q_2}$ in diffraction pattern illustrate phase-separated state in the crystal at room temperature. (b) The stripe and superstructure spots disappear at temperature 530K, showing the high temperature homogeneous phase without vacancies ordering. (c) The crystal recovered into the phase-separated state in the cooling run.

Ⅱ. **Simulation of the PSP using phase filed theory**

We herein present the theoretical data obtained from a 2-dimensional Monte Carlo simulation using the conventional phase-field approach to investigate the microstructure evolution in the PSP process of the homogeneous high-temperature solution. Since only morphological patterns in **a-b** plane are considered, the free energy F employed in our simulation is similar with that used in previous literatures Ref. [2], where the elastic energy between the two phases are added in the free energy of time-dependent Cahn-Hilliard equation.

$$F = \int dxdy [f(c) + \tfrac{1}{2}(\nabla c)^2 + B_{e1}]$$

Where c(**r**, t) represents the local concentration of the solid solution at sit **r** and time t, f(c) = c(1-c)(1+c) is the double-well potential function for local free-energy density which could produce equilibrium composition of two phases in the free diffusion process, and $B_{el}$ is the elastic energy related to the lattice misfit between $\mathbf{q_1}$ and $\mathbf{q_2}$ phase at the interface. The softest directions are set as [1-10] and [110] considering the strain anisotropy introduced to the system by Fe-ordering in the Fe-Se squared lattices, and the domains are rectangular stripes aligned along the softest directions in **a-b** plane. The elastic energy is related with the composition and lattice mismatch between Fe-rich and poor phases, and the domains are rectangular stripes aligned along the [110] and [1-10] directions.

Our simulations were performed on a square area discretized by a lattice of 256×256 grid points with periodic boundary conditions. The discretizing grid size Δx is chosen to be 1.0 and the time step Δt is 0.01. The overall scaled composition variable is zero, which corresponds to a critical composition. The values of composition c(r, $t_0$) are uniformly distributed between 0.1 and -0.1 as the initial configuration, corresponding to the high-temperature homogeneous state where the composition deviation from the average value is only caused by fluctuations. c=±1 represents for the equilibrium compositions of $q_1$ and $q_2$ phase. The fractions of $q_1$ and $q_2$ phases in our simulation are regulated by the critical composition.

### III. Impact of thermal treatment on microstructure evolution for $K_{0.8}Fe_{1.75}Se_2$

Our systematic investigations on $K_{0.8}Fe_{1.75}Se_2$ SC sample revealed that morphological pattern and microstructure feature could be changed visibly by tuning the annealing conditions, further affecting the superconductivity. Samples with nominal composition $K_{0.8}Fe_{1.75}Se_2$ annealed for different time from 600K to RT were used. As shown in the SEM images for $K_{0.8}Fe_{1.75}Se_2$ of Fig. S3(a-d), stripes with different wavelengths between 20nm and 2μm were observed. The simulated images presented in Fig. S3(e-h) are consistent with the experimental ones. Resistivity measurements of the samples reveal that the SC transition temperatures increase as the annealing time increases, but the superconductivity is suppressed when the annealing time is prolonged further.

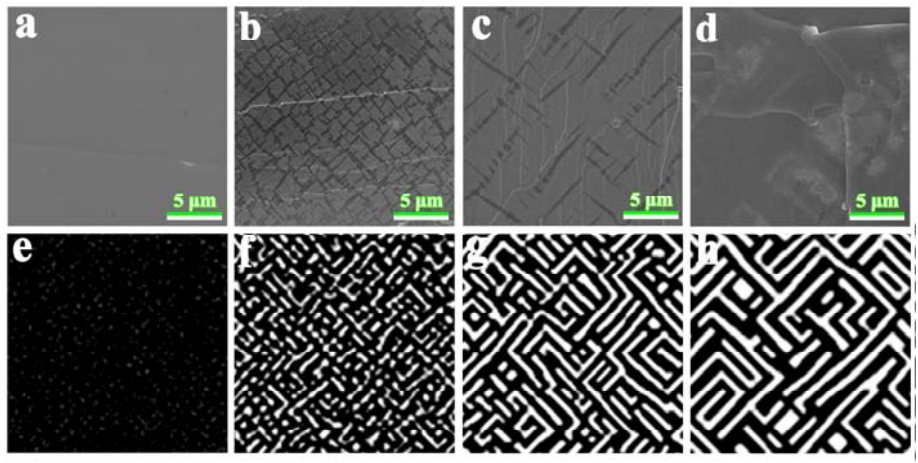

**Fig. S3** SEM images of **a-b** plane for $K_{0.8}Fe_{1.75}Se_2$ annealed at 600K using different heat treatment methods illustrate the time development of SPS process: (a) ice water cooling, (b) liquid nitrogen cooling, (c) furnace cooling (as grown) and (d) controlled cooling (72h to RT). Simulated 2D morphological patterns during SPS at (e)

t=50, (f) t=600, (g) t=3000 and (h) t=15000.

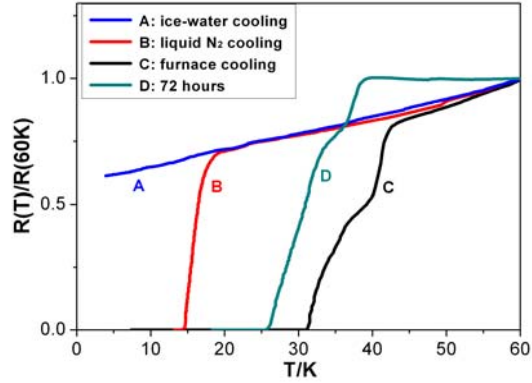

**Fig. S4** In plane temperature-dependent resistivity profiles of $K_{0.8}Fe_{1.75}Se_2$ samples annealed with different thermal process.

## Ⅳ. Microstructures in Fe-rich domains

Our SEM investigations on $K_{0.8}Fe_{1.6+x}Se_2$ SC samples with higher iron content illustrate that the $q_2$ SC phase forms well-developed discontinuous precipitation morphology surrounded by the AFM $K_2Fe_4Se_5$ phase in the stripe area, and these domains also align themselves along [110] and [1-10] direction in **a-b** plane, resulting different kinds of stripe patterns as typical shown in Fig. S5. The system evolves in such a way in order to in order to compensate for the increase of interface energy and minimize the total energy.

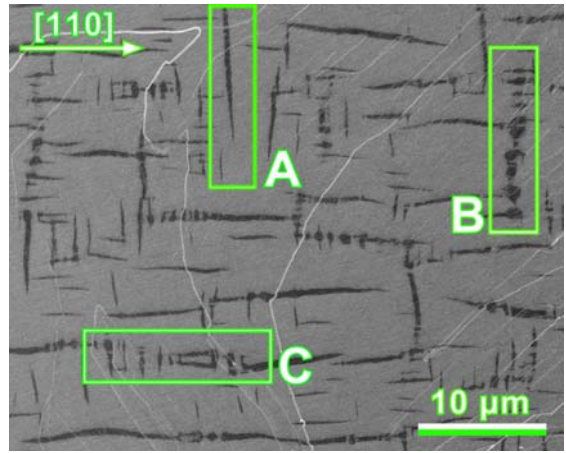

**Fig. S5** SEM image for $K_{0.8}Fe_{1.75}Se_2$ samples of **a-b** plane shows the stripe patterns aligned along [100] and [1-10] directions. Three kinds of stripes with different morphology are indicated: (A) long stripes with small relief on each side, (B) rectangular domains of $q_2$ phase and (c) open squared domains of $q_2$ phase.

## Ⅴ. Defect structure in the Fe-rich domains

HRTEM studies on $K_{0.8}Fe_{1.6+x}Se_2$ samples reveal two typical defect structures in the stripe area as shown in Fig. **S6** and Fig. **S7**. STEM image of $K_{0.8}Fe_{1.75}Se_2$ in Fig. S6(a) shows the complex contrast in the stripe area, where the spindle area with the dark contrast is caused by the low atomic density. A new squeezed phase with **c**-axis lattice parameter 11.2Å is observed between the spindle area, which could be induced by releasing stress during the formation of $\mathbf{q_2}$ phase. It is also suggested that the squeezed phase maybe has a relation with the SC transition at 43K. Intergrowth of $\mathbf{q_1}$ phase, $\mathbf{q_2}$ phase and the squeezed phase along the **c**-axis zone direction is shown in Fig. S7, which could be in correlation with the suppression of $T_c$.

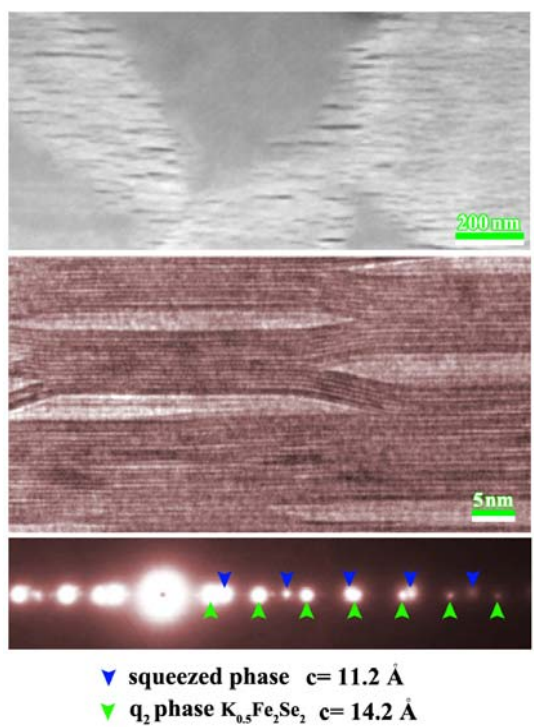

▼ squeezed phase c= 11.2 Å
▼ $q_2$ phase $K_{0.5}Fe_2Se_2$ c= 14.2 Å

Fig. S6 (a) STEM image for $K_{0.8}Fe_{1.75}Se_2$ sample along [310] zone-axis direction. (b) HRTEM image in the stripe area with a squeezed phase between the spindle areas. (c) Selected area electron diffraction with electron beam perpendicular to c-axis.

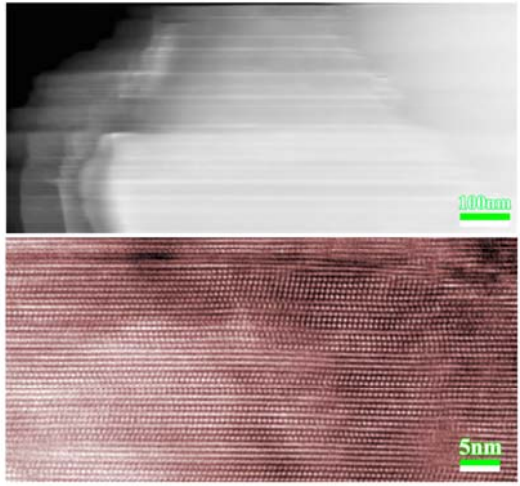

Fig. S7 (a) STEM image for $K_{0.8}Fe_{1.75}Se_2$ sample along [310] zone-axis direction. (b) HRTEM image of (310) plane show the intergrowth of $q_1$ phase, $q_2$ phase and the squeezed phase.

## VI. SC volume fraction estimated based on the Archimedean solid model.

The hollow truncated octahedron can be formed from a large truncated octahedron via emptied by removing a smaller one. The truncated octahedron is built by eight half tetragonals. The volume of the large half tetragonal is given by the volume of large tetragonal with side length $a \times a \times c$

$$V_{Lht} = \frac{1}{2}a^2 c$$

The volume of the small half tetragonal is obtained by the volume of small tetragonal with side length $a\delta \times a\delta \times c$

$$V_{Sht} = \frac{1}{2}a^2 \delta^2 c$$

where $\delta$ could be obtained from the width of the stripes ($W_d$) and the width of corresponding observed area ($W$) in the SEM image, i.e. $\delta = 1 - \frac{W_d}{W}$.

Based on the Archimedean solid model, the SC volume fraction could be represented by the ratio of the hollow truncated octahedron and the large tetragonal

$$f_{sc} = \frac{V_{Lht} - V_{Sht}}{V_t} = \frac{1}{2}(1-\delta^2)$$